\documentclass[a4paper,10ptd]{revtex4}

\usepackage{amsmath}
\usepackage{amssymb}
\usepackage{graphics}

\newcommand{\de}[1]{\left( #1 \right)}

\newcommand{\DE}[1]{\left\{ #1 \right\}}

\newcommand{\ket}[1]{\left| #1 \right\rangle}

\newcommand{\ie}{{\it{i.e.}}: }

\newtheorem{theorem}{Theorem}
\newenvironment{proof}{{\newline \bf{Proof:}}}{{\fbox{ }}}

\begin{document}

\title{Entanglement in single particle systems}
\author{M. O. Terra Cunha, J. A. Dunningham, and V. Vedral} 
\address{
School of Physics and Astronomy, University of Leeds, Leeds LS2 9JT, United Kingdom.}

\begin{abstract}
We address some of the most commonly raised questions about entanglement, especially with regard to so-called occupation number entanglement. To answer unambiguously whether entanglement can exist in a one-atom delocalized state, we propose an experiment capable of showing violations of Bell's inequality using only this state and local operations. As a byproduct, this experiment suggests a means of creating an entangled state of two different chemical species. By comparison with a massless system, we argue that there should be no fundamental objection to such a superposition and its creation may be within reach of present technology.
\end{abstract}

\maketitle

\section{What is Entanglement?}
\label{What}
Ever since entanglement has been characterized as a resource, people have asked questions like {\emph{is there entanglement in such a state/system?}} or even more subtley {\emph{how much entanglement is there in such a state/system?}}. In this paper, we want to emphasise that, in isolation, these questions are meaningless. They presuppose some important details which, if left unstated, can give rise to considerable misunderstanding.

Bohr pointed out that any discussion of quantum mechanics must be contextual, in the sense that a concrete physical situation should be described, including the measuring apparatus \cite{Bohr}. In the same way, even the most abstract discussions of entanglement must clearly state which subsystems are being considered, since it is  between these subsystems that entanglement may or may not appear. Identical states can exhibit entanglement between certain subsystems in one description and yet not in another. A good example is provided by the simple and well-studied system of a Hydrogen atom \cite{QM}. To diagonalize the Hydrogen atom Hamiltonian, the first step is to change coordinates from proton and electron to the centre of mass and relative position of the particles. The centre of mass is free, and the problem is reduced to the relative separation, which is subject to a central potential. One then finds the eigenstates of the relative particle and, due to the $10^3$ ratio between the masses of the proton and electron, it is a good approximation to call these the {\emph{electronic states}} of the atom. Consequently the ground state of a Hydrogen atom is a tensor product of the ground state of the free centre of mass (i.e. a plane wave of zero momentum) and the ground state of the relative particle (i.e. the spherical $1s$ orbital). There is no entanglement in this description. However, if one returns to a description in terms of the proton and the electron, the pure state that results is non-factorizable and hence entangled \cite{Pisa}. It is true that the large mass assymetry implies that the degree of entanglement will be small. Nevertheless, one could instead consider the positronium system in the same context and the result would be that the positron and electron were as entangled as the particles in the original EPR state\cite{EPR}.

Another physically appealing example of such a situation is a system of interacting harmonic oscillators. The first step to solve this problem is to transform to normal modes. In doing so, the eigenstates of the system are tensor products of the eigenstates of the normal modes and, as such, there is no entanglement at all between these modes. However, entanglement may be seen in a description of the original oscillators modes. Similarly, the vibrational ground state of a lattice will be a direct product of ground states of each of its phonon modes, but this same state shows entanglement among the constituent ionic cores \cite{Cond}. 

The essential point here is that entanglement is not an {\emph{absolute}} property of quantum states. Entanglement is a property of a quantum state {\emph{relative}} to a given set of subsystems. Even the quantum information community's favored system of a pair of qubits has been shown to have an infinity of {\emph{Tensor Product Structures}} \cite{TPS}, which means that the same two qubit state can be entangled or not, depending on the structure chosen. This last example illustrates another common misconception. The other three examples we gave were based on Hamiltonians that coupled the subsystems, however one must avoid thinking that direct interaction is necessary for the appearance of entanglement \cite{Swap}. It is just one easy and natural way of (generically) creating entanglement among the subsystems.

In the appendix we state and prove a Theorem which says that any given pure state of a state space, which can be written as a composite system, can be viewed as an unentangled state. If one also remembers that a generic pure state decomposed into well-defined subsystems, is not factorizable, one can understand why a question like ``is there entanglement in such a state?'' is meaningless.

In the next section we consider an even more controversial point \cite{VEnk}: is it possible to have entanglement in a one photon state?

\section[Particles vs Modes]{Entanglement of Particles vs Entanglement of Modes}
The vast majority of the discussions about entanglement start with a well defined system of particles such as two photons, or three atoms, or one atom and one photon. At the same time, the quantum information community would like to consider entanglement to be a fundamental aspect of quantum mechanics that transcends any specific realization.

If we want to increase the breadth of applicability of entanglement we should think in terms of fields, which are a fundamental description of nature. Particles are only a manifestation of certain special configurations of quantum fields. If entanglement is to be considered a fundamental property of nature, and even a resource to be understood and applied, one would like to understand entangled fields.

To be more specific, let us discuss what happens to a photon when it reaches a beam splitter. If we consider just one spatial mode across the beam splitter and another reflected, in ocupation number notation this state is (normalization omitted)
\begin{equation}
\ket{10} + \ket{01},
\label{01}
\end{equation}
which is just a Bell state for the two mode system. But which are the two subsystems to be entangled? In this example, the two spatial modes. It is the state of the two spatial modes that is non-factorizable.

One will have much trouble in denying the entanglement in this one photon state. A natural question is, {\emph{can one make a Bell's inequality measurement in this system?}}, and the answer is yes. An ``easy'' way is to resonantly couple each mode with a two level atom for a time corresponding to a $\pi$ pulse, \ie the photon is absorbed and the corresponding atom excited. In this case, the mode state will always end up in the vacuum, while the atoms assume the Bell state
\begin{equation}
\ket{eg} + \ket{ge}.
\label{eg}
\end{equation}
This state is now well-suited to carrying out experiments to test Bell's inequality. Furthermore, one can say that this ``transfer of state'' is part of the detecting apparatus, and that the Bell's inequality experiment is performed directly on the two-mode one-photon system.

In this sense we conclude that another usual fallacy is saying that {\emph{``entanglement is a property of many-particle systems''}}. It is, however, correct to say that {\emph{entanglement is a property of composite systems}}, i.e. systems that have more than one subsystem.

\section{Entanglement and Locality}
Another common misconception is to consider entanglement and non-locality as one and the same issue, since locality naturally implies position distinguishability, which is not a condition for entanglement. To clarify this point, we will describe a Bell measurement showing entanglement, which can not be considered as a manifestation of non-locality. In fact, it is just another system described by the state \eqref{01}.

The origin of this (incorrect) generalization is natural: non-locality is a non-classical manifestation of quantum systems and a large class of applications of entanglement deal with spatially separated laboratories, usually occupied by Alice and Bob. In such contexts, it seems natural to discuss non-locality. This is the origin of the idea of {\emph{Local Operations}}, and generally, it is in this sense that the quantum information community talks about non-locality, rather than the sense implied by Relativistic Theory.

Let us now consider a photon linearly polarized at $45^o$. In the occupation number notation and with respect to horizontal and vertical polarization modes, such a photon is described by the state \eqref{01}. Again, it is an entangled state of these two modes, and it is again a good example of how different descriptions of one and the same system can point to diferent answers to the question {\emph{is it entangled?}} This state is factorizable if one uses the modes with crossed polarizations, \ie $45^o$ and $135^o$.

This leads us to the question of {\emph{Can we make a Bell measurement?}} for this system, or in a wider context, {\emph{can we measure an entanglement witness?}} Again, the answer is yes. The simplest strategy involves using a polarizing beam splitter to spatially separate the two modes and then the same strategy as before can be used. One can even say that the beam splitter creates non-locality, since the state after the beam splitter is the same as the one discussed in the previous section. However, as in the previous section, all this can be considered as part of the measuring apparatus and, in this case, we measure entanglement at just one position in space, to which it does not seem fair to talk about non-locality. 

In this example, we said that the beam splitter generates non-locality, but in accordance with section \ref{What}, one should not say that it creates entanglement. A beam splitter is an example of a linear mode-converter. It can naturally be interpreted as a change in the Tensor Product Structure, more specifically, as a change in which operators are directly spatially measured. It has been shown that it can even convert one kind of non-classicality (squeezed states) into another kind of non-classicality (entanglement) \cite{DGCZ}. 

\section{An experiment showing entanglement of massive fields}
Up until this point, we have discussed entanglement of massless fields, involving one photon distributed over two modes that then become entangled. We would now like to consider a similar situation involving one atom distributed over two spatially distinct locations. Such a state must be entangled with respect to the usual local tensor product structure of particle location. In particular, we would like to consider the question:  {\emph{``is there any superselection rule which prevents massive one-particle entanglement?''}}

The physical set-up begins with a single atom that is in a coherent superposition of being in two atomic traps, $T_1$ and $T_2$. This can be described by the state
\begin{equation}
\ket{A0} + \ket{0A},
\label{A0}
\end{equation}
where $A$ means that there is an atom and $0$ means no atom at all. The ordering refers to traps $1$ and $2$. This state is a direct analog of Eqs. \eqref{01} and \eqref{eg}.

\begin{figure}
	\includegraphics{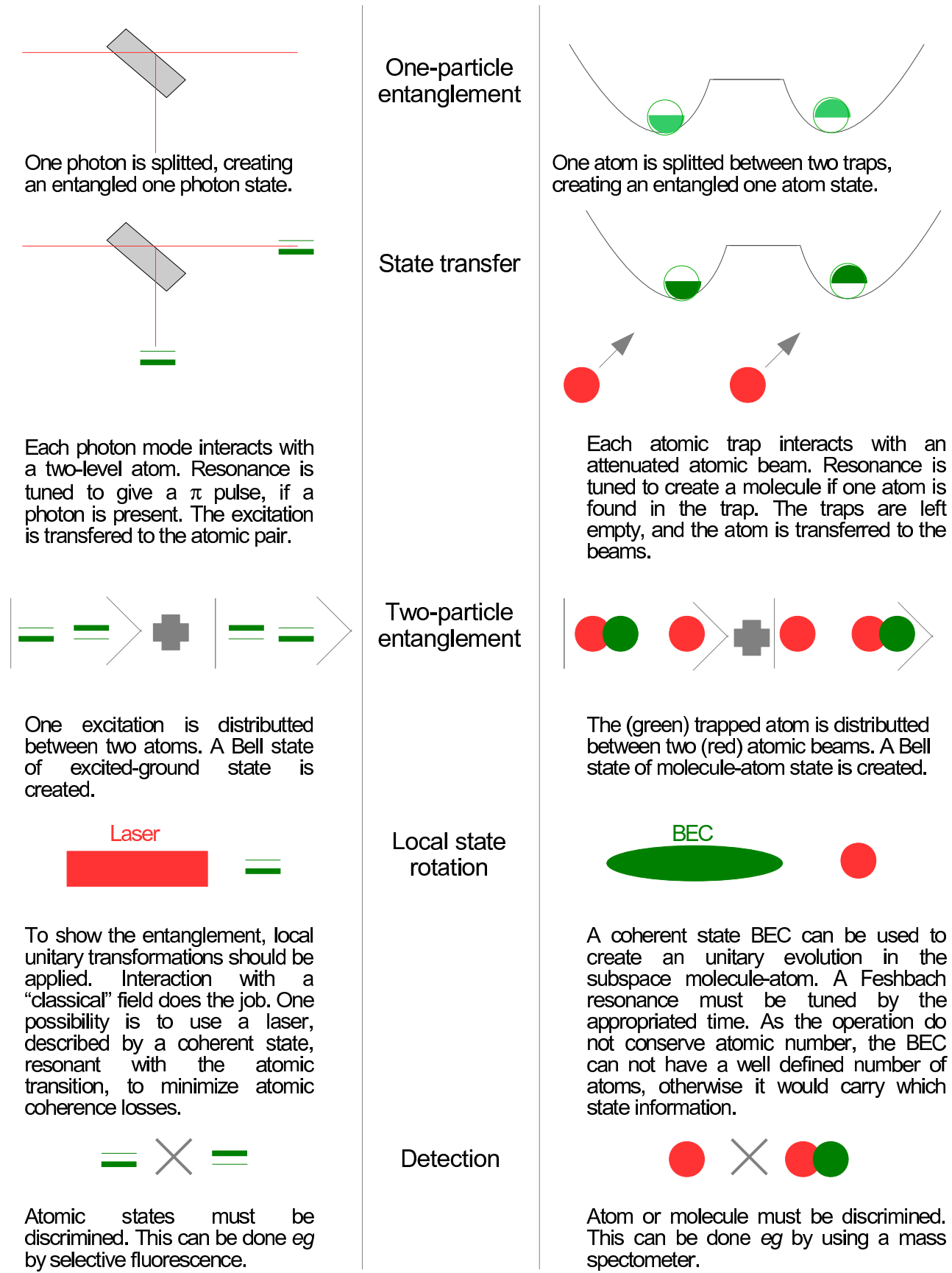}
	\caption{Schematic description of two experiments to find evidence for single particle entanglement, one with a single photon, the other with a single atom.}
	\label{fig}
\end{figure}
As in the photonic case (see Fig. \ref{fig}), it is important that something else `drags' this atom for us to make direct measurements that confirm its entanglement. Two atomic beams $B_1$ and $B_2$ must be focused on the traps with controlling parameters (such as classical magnetic fields which tune a Feshbach resonance \cite{Feshbach}) choosen in a way that if there were an atom $A$ in trap $T_i$ then it would be captured by the beam forming a molecule $AB$ at the beam $i$, while trap $T_i$ would be left empty. For simplicity of the argument, let us consider just one atom $B_i$ in each beam. This time evolution (similar to the $\pi$ pulse described above) can thus be described as
\begin{equation}
\DE{\ket{A0}+\ket{0A}}\ket{B_1,B_2} \mapsto \ket{00}\DE{\ket{B_1A,B_2} + \ket{B_1,B_2A}}.
\label{assoc}
\end{equation}
So far we have not specified the $B_i$ atoms. They can be equal to each other, or not. Also, they can be equal to the atom $A$ or not. For the detection mechanism we have devised, it will be important that they are not equal to $A$. However, this is not a fundamental problem.

At this stage of the scheme, we have an entangled state of two chemically distinct specimens: atom and molecule. Now we have to test this entanglement, for example through a witness like a Bell inequality, or by using it to perform a classically inadmissable operation such as teleportation. In the remainder of this analysis we will neglect the state of traps $T_1$ and $T_2$, since they have factored out from the flying modes.

One straightforward operation that we could perform on a state like
\begin{equation}
\ket{B_1A,B_2} + \ket{B_1,B_2A}
\label{mol-at}
\end{equation}  
is just to measure whether there is an atom or a molecule in each spatial position. For this  a mass spectrometer can be used. Another is to impinge a relative phase, $\phi$, between them, creating the states of the form
\begin{equation}
\ket{B_1A,B_2} + e^{i\phi} \ket{B_1,B_2A}.
\label{mol-atphi}
\end{equation}  
To achieve this, one only needs to apply a DC electric field in one of the spatial modes, since the Stark shift of the atom and the molecule should be different.

However, these two operations alone do not test entanglement. In a qubit language, we are only measuring in the computational basis, and applying a conditional phase shift to a state like $\ket{\Psi_+}$. We need to implement other local operations. To test a Bell inequality, we must be able to change basis. It is important to stress what {\emph{local}} means here: it means that beam $1$ and beam $2$ must be addressed individually. The transformations we need act in each beam separately, and do the following:
\begin{subequations}
\label{at-molpiover2}
\begin{eqnarray}
\ket{BA} & \mapsto & \ket{BA} - \ket{B};\\
\ket{B} & \mapsto & \ket{BA} + \ket{B}.
\end{eqnarray}
\end{subequations}
Clearly, such transformations do not conserve the number of $A$ atoms locally, and can not be implemented without extra $A$ atoms. But they can take place in other traps with other $A$ atoms in each trap, to which each beam is again focused. Let us initially pretend to have one $A$ atom in each of these additional traps, just to discuss the situation. We need to find a Feshbach resonance so that in the presence of the appropriate field, the system evolves cyclically like a two level system: $A + B \mapsto AB \mapsto A + B$, with period $t$. If one waits only $\frac{t}{4}$, one will implement
\begin{subequations}
\begin{eqnarray}
\ket{B,A} & \mapsto & \ket{BA} + \ket{B,A};\\
\ket{BA,A} & \mapsto & \ket{BA,A} - \ket{B,A,A}.
\end{eqnarray}
\end{subequations}
Such operation, followed by an atom-or-molecule detector almost implements our ``reading in a different basis'' scheme. However, the number of $A$ atoms have ``which path'' information, and for this to work, it would be necessary to have a physical impossibility (and not a technical one) of knowing the number of $A$ atoms remaining in the traps. Alternatively, one can say that for this situation there is a superselection rule for the number of $A$ atoms \cite{VC}, which must be surpassed for the implementation of the time evolution \eqref{at-molpiover2}.

Despite these apparent problems, it can still be done. If we prepare a condensate in a state like a coherent state (which would also affect the interaction time, but this can be taken into account), then there would be essentially no difference between the original state in the trap and the state with one added atom. One should remember at this time that in order to apply a Ramsey pulse in the optical regime one usually uses a laser, and if one tries to do it with a Fock state it would simply not work, since the atom and the field would become entangled, destroying the coherence of the atomic state. This optical analogy is emphasized in Fig.~\ref{fig}.

After this change of basis, one only needs to discriminate between atom and molecule in each spatial mode to complete a Bell measurement for a one-atom entangled state.
Of course, once again we are free to consider everything that happens after the state \eqref{mol-at} has been created as just being part of the ``detector''.

\section{Concluding remarks}

In this paper we have argued that the natural arena for discussing fundamental questions of entanglement is quantum field theory. Particles are effective concepts, and effective theories are welcome whenever they can be applied. In this sense, the usual scenario of a well-defined number of multiple particles and subsystems is useful. However, there are certain contexts when this must be abandoned if we are to treat entanglement in  a truly general sense. One consequence of this is the possibility of single particle entanglement.

One-particle entanglement is as good as two-particle entanglement with respect to applications. A one-photon or one-atom state can be used to teleport a qubit, provided that it is delocalized. Finally, we would like to emphasise that, despite both being manifestations of non-classicality, entanglement and non-locality are not synonyms.

\appendix
\section{Statement and Proof of a Theorem}
\begin{theorem}
Given a state vector $\ket{\psi}$ in a finite dimensional state space $\mathcal{H}$ with nonprime dimension $d = mn$, there exists a tensor product structure ${\mathcal{H}} \equiv {\mathcal{V}}^m \otimes {\mathcal{W}}^n$ with respect to which  $\ket{\psi}$ is factorizable.
\end{theorem}
This theorem must be compared o two other results on linear algebra, with important applications on quantum mechanics. The first one is that given any state vector there exists an orthonormal basis including such a vector. It is indeed in the core of the proof of the above Theorem. The other is the so called Schmidt decomposition, in which given a state vector $\ket{\psi}$ and a bipartite tensor product structure ${\mathcal{H}} \equiv {\mathcal{U}} \otimes {\mathcal{V}}$, one can choose orthonormal basis $\DE{\ket{u_i}}$ and $\DE{\ket{v_j}}$ on the factors and write $\ket{\psi} = \sum _k \lambda_k \ket{u_k}\otimes \ket{v_k}$.  The similarity is that one must remember that the Theorem begins with the state $\ket{\psi}$ given. The tensor product structure will be adapted for this state.
\begin{proof}
One must remember that isomorphic vector spaces can be identified. So just take two vector spaces ${\mathcal{V}}'$ and ${\mathcal{W}}'$ and choose orthonormal basis $\DE{\ket{v_i}}$ and $\DE{\ket{w_j}}$ for them. Choose an ordered orthonormal basis $\DE{\ket{h_k}}$ for ${\mathcal{H}}$ with $\ket{\psi}$ as its first element. Set the isomorfism ${\mathcal{V}}'\otimes {\mathcal{W}}' \rightarrow {\mathcal{H}}$ given by $\ket{v_i}\otimes \ket{w_j} \mapsto \ket{h_{\de{i-1}\times n + j}}$. This isomorfism induces a Tensor Product Structure in ${\mathcal{H}}$ and with respect to it, $\ket{\psi}$ is factorized.  
\end{proof}

\end{document}